\documentclass[11pt]{article}

\usepackage[final]{acl}

\usepackage{times}
\usepackage{latexsym}

\usepackage[T1]{fontenc}

\usepackage[utf8]{inputenc}

\usepackage{microtype}

\usepackage{inconsolata}

\usepackage{graphicx}

\usepackage{amsmath}   
\usepackage{amssymb}   

\usepackage{listings}  

\usepackage{booktabs}  
\usepackage{multirow}  
\usepackage{tabularx}  
\usepackage{array}     

\newcolumntype{Y}{>{\raggedright\arraybackslash}X}
\newcolumntype{P}[1]{>{\raggedright\arraybackslash}p{#1}}

\usepackage{tikz}      
\usepackage{xcolor}    

\newcommand{\fsymbol}{%
  \tikz[baseline=(char.base)]{
    \node[shape=circle, draw=green!60!black, fill=green!20,
          inner sep=0.2pt, minimum size=1.1em,
          text=green!50!black, font=\scriptsize\bfseries] (char) {F};
  }%
}

\newcommand{\psymbol}{%
  \tikz[baseline=(char.base)]{
    \node[shape=circle, draw=red!70!black, fill=red!15,
          inner sep=0.2pt, minimum size=1.1em,
          text=red!60!black, font=\scriptsize\bfseries] (char) {P};
  }%
}

\title{DIALECTIC: A Multi-Agent System for Startup Evaluation}

\author{
    \textbf{Jae Yoon Bae\textsuperscript{1,2,*}},
    \textbf{Simon Malberg\textsuperscript{1,*}},
    \textbf{Joyce Galang\textsuperscript{1,3,*}},
    \\
    \textbf{Andre Retterath\textsuperscript{2}},
    \textbf{Georg Groh\textsuperscript{1}}
    \\
    \\
    \textsuperscript{1}Technical University of Munich,
    \textsuperscript{2}Earlybird Venture Capital,
    \textsuperscript{3}UVC Partners
    \\
    \textsuperscript{*}These authors contributed equally to this work.
    \\
    \small{
        \textbf{Correspondence:} \href{mailto:jaeyoonbae99@gmail.com}{jaeyoonbae99@gmail.com}
    }
    \\
    \small{
        \textbf{Code:} \href{https://github.com/pantageepapa/DIALECTIC}{github.com/pantageepapa/DIALECTIC}
    }
}

\begin{document}
\maketitle
\begin{abstract}
Venture capital (VC) investors face a large number of investment opportunities but only invest in few of these, with even fewer ending up successful. Early-stage screening of opportunities is often limited by investor bandwidth, demanding tradeoffs between evaluation diligence and number of opportunities assessed. To ease this tradeoff, we introduce DIALECTIC, an LLM-based multi-agent system for startup evaluation. DIALECTIC first gathers factual knowledge about a startup and organizes these facts into a hierarchical question tree. It then synthesizes the facts into natural-language arguments for and against an investment and iteratively critiques and refines these arguments through a simulated debate, which surfaces only the most convincing arguments. Our system also produces numeric decision scores that allow investors to rank and thus efficiently prioritize opportunities. We evaluate DIALECTIC through backtesting on real investment opportunities aggregated from five VC funds, showing that DIALECTIC matches the precision of human VCs in predicting startup success.
\end{abstract}

\section{Introduction}

The global venture capital (VC) industry is expanding rapidly alongside intensified competition for attractive deals. The market is projected to grow from USD~337~billion in 2024 to USD~1.46~trillion by 2033, a compound annual growth rate of 17.6\% \citep{IMARC2024}. Entrepreneurial activity has also surged, with annual U.S. business formations rising from 3.5~million in 2019 to 5.2~million in 2024, an increase of nearly 40\% \citep{USCensus2025}. Traditional VC decision-making processes are challenged in this setting. Investors face high time pressure and information overload, both associated with suboptimal decisions \cite{zacharakis2001natureofinfo}. These conditions have increased interest in computational approaches for scalable investment evaluation.

Among these approaches, machine learning methods have emerged as a promising direction. Prior studies demonstrate strong predictive performance that, in some cases, even surpasses human investors \cite{antretter2019predicting,retterath2020humanvscomp,zacharakis2001natureofinfo,arroyo2019assessment,dellermann2021findingunicornpredictingearly,sharchilev2018startupsuccess}. Yet, these non-iterative models diverge from how investment decisions are formed by human VCs. In practice, conviction emerges through iterative hypothesis formation, challenge, and refinement as new information appears \cite{chong2014constructng}.

Recent advances in \textit{large language model} (LLM) orchestration tools enable iterative and interpretable reasoning. Frameworks such as \textit{LangChain} \cite{LangChain} support the decomposition of complex tasks, the generation of intermediate conclusions, and the iterative refinement of responses while making the underlying logic explicit. They allow multi-step reasoning and dialectical interaction, a setup in which LLMs can articulate arguments, generate counterpoints, and produce transparent reasoning traces.

This paper introduces \textit{\textbf{D}ecision \textbf{I}teration with \textbf{A}rgument-\textbf{L}evel \textbf{E}vidence and \textbf{C}ounter-\textbf{T}hinking for \textbf{I}nvestment \textbf{C}onclusions} (\textbf{DIALECTIC}), an LLM-based system that models iterative and argumentative elements of venture evaluation. Our system draws on principles of dialectical reasoning, an approach shown to be effective for complex, unstructured problems that benefit from structured confrontation of differing perspectives \citep{jarupathirun2007dialectic}. The contributions of this work are:

\begin{itemize}
    \item {A structured LLM reasoning system that models how investors build and refine investment theses through argumentation.}
    \item {An empirical evaluation demonstrating predictive performance in venture screening.}
\end{itemize}

Overall, the proposed system brings data-driven VC methods closer to industry practice. Furthermore, it enables the process of iterative argumentation in early-stage screening, which has traditionally been restricted to later stages of the decision funnel due to limited investor bandwidth. This shift allows investors to apply iterative reasoning earlier in the process, improving both diligence quality and screening efficiency.

\section{Related Work}
\label{sec:related_work}

Prior studies propose different machine learning approaches to predict startup outcomes, drawing on public data sources such as \textit{Crunchbase} \citep{arroyo2019assessment, zbikowski2021biasfree, retterath2020humanvscomp}, \textit{Twitter} \citep{antretter2019predicting}, web data \citep{sharchilev2018startupsuccess}, and \textit{Google Search} \citep{gavrilenko2023improvingstartupsuccess}, and often reporting promising prediction accuracy (see Table \ref{tab:related_work_overview} in the Appendix for an overview). Most studies trained gradient tree boosting models (e.g., \textit{XGBoost}) \citep{corea2021hacking, arroyo2019assessment, zbikowski2021biasfree, retterath2020humanvscomp} and interpreted predictions using feature-importance rankings with features such as geography, industry, or founder background \citep{zbikowski2021biasfree, sharchilev2018startupsuccess, gavrilenko2023improvingstartupsuccess}.

Some newer studies have used LLMs to extract structured features or embeddings from unstructured data, while still resorting to machine learning models such as \textit{XGBoost} for prediction \citep{ozince2024vcfounder, maarouf2025fused}. \citet{xiong2023founderideafit} used LLMs to assess founder-idea fit, also providing pro and contra arguments for interpretability. In follow-up work \citep{xiong2024gptree}, they focus on extracting traits associated with successful entrepreneurs. Both studies look at individual founders rather than startup companies.

Beyond VC, LLM-based decision-making frameworks have been proposed for fields such as business or finance. \textit{DeLLMa} combines LLMs with decision-theoretic reasoning \citep{liu2025dellma}, while \textit{STRUX} extracts facts from companies' earnings calls and produces weighted pro and contra aspects for buy or sell decisions \citep{lu2025strux}.

A promising approach to improving LLM reasoning is the introduction of multi-agent systems \citep{han2024multiagent}. Instead of relying on a single model, several LLMs interact through collaboration, debate, or specialization. In adversarial or collaborative debating, agents defend opposing stances and a separate judge model or heuristic evaluates the quality of their arguments \citep{chan2023chateval, liang2024multiagentdebate}.

\section{DIALECTIC}
\label{sec:method}

\begin{figure*}
    \centering
    \includegraphics[width=1\linewidth]{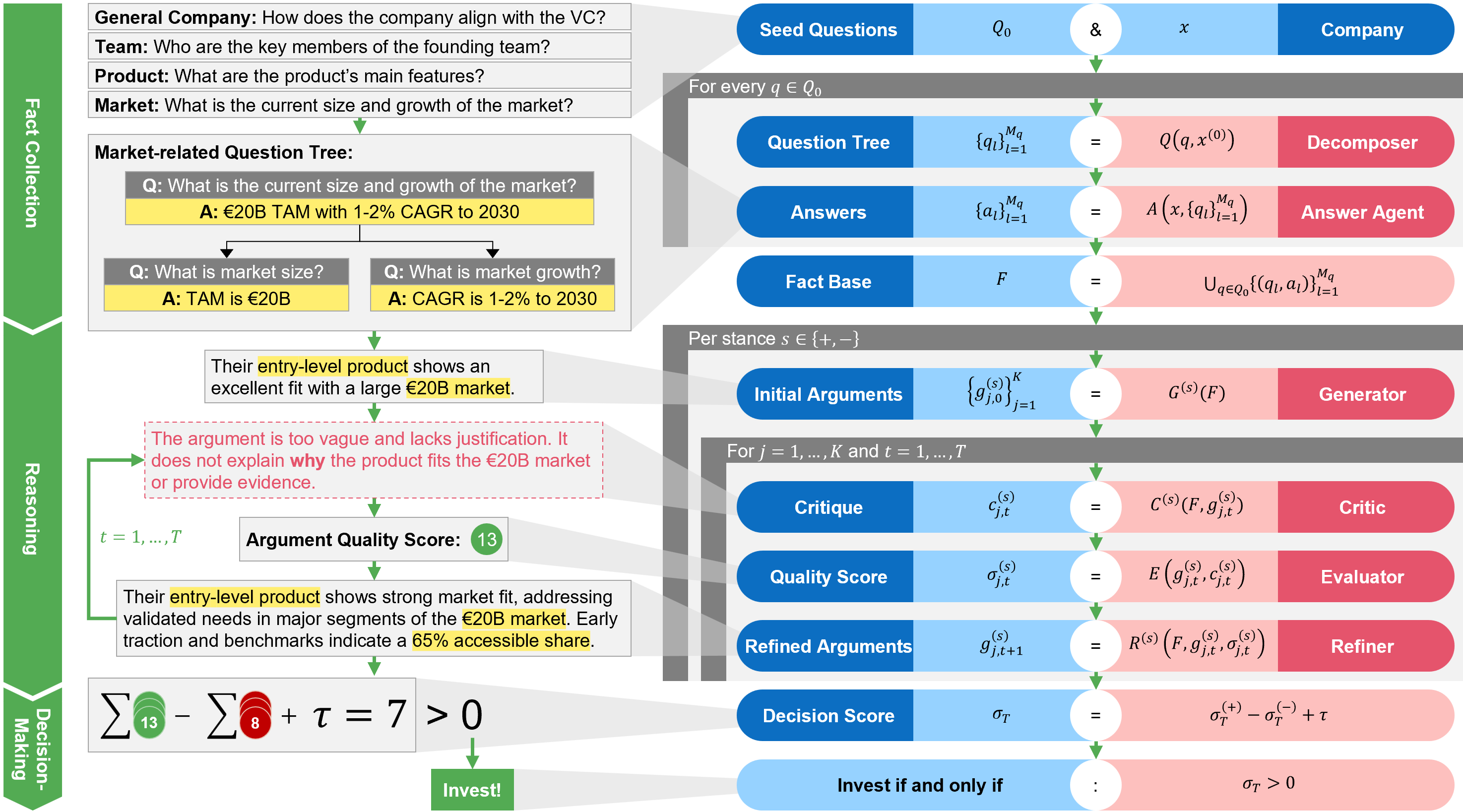}
    \caption{Overview of the DIALECTIC method. The right side shows the flow of operations. Agents are shown in red, agent inputs/outputs are shown in blue, and loops are shown in green. The left side illustrates the key outputs of the agents.}
    \label{fig:method}
\end{figure*}

DIALECTIC is inspired by how real VC investors make investment decisions. They collect information about a startup, form narrative investment hypotheses, and refine these hypotheses through debate with other VCs until making a decision. DIALECTIC proceeds in three phases: \textbf{fact collection}, \textbf{reasoning}, and \textbf{decision-making}. During fact collection, DIALECTIC gathers factual knowledge about a company and organizes these facts hierarchically in a question tree. In the reasoning phase, it synthesizes raw facts into arguments pro and contra an investment, which it iteratively self-critiques, evaluates, and refines, letting only the best arguments survive. Finally, it makes a decision based on a comparison of the best pro and contra arguments; see Figure \ref{fig:method} for an illustration.

In the following, we formally introduce DIALECTIC. Let $X=\{x_i\}_{i=1}^N$ be the set of investable companies, each described by multiple features $x_i^{(d)},d=1,...,D$. 
The goal is to predict the ground truth label $y_i \in \{\text{successful},\text{unsuccessful}\}$ signaling whether the company will be successful and should be invested in or not.

\subsection{Fact Collection Phase}
We denote the universe of natural-language questions as  $\mathcal{Q}$, the universe of natural-language answers as $\mathcal{A}$, and the set of industries as $I$. For a given company 
$x$ in industry $x^{(0)} \in I$, we start by providing DIALECTIC with a set of \textbf{seed questions} $Q_0 \subset \mathcal{Q}$. Specifically, we ask four questions about the general company, team, product, and market (see Appendix \ref{sec:appendix_seed_questions} for details) to cover the main aspects typically considered by VC investors \citep{retterath2020humanvscomp}. Inspired by \textit{ProbTree} \citep{cao2023probtree} and \textit{Socratic Questioning} \citep{qi2023socraticquestioning}, we define two LLM-based agent operations:

\begin{itemize}
    \item The \textbf{decomposer} $Q:Q_0\times I \rightarrow \mathcal{Q}^*$ takes a seed question $q \in Q_0$ and hierarchically decomposes it into a finite set of $M_q$ sub-questions relevant in the industry, thus creating an industry-specific \textbf{question tree}\footnote{For simplicity, we do not explicitly model the hierarchical organization of questions in our notation but represent question trees as simple sets. However, our code implementation preserves the full hierarchy.} $\{q_l\}_{l=1}^{M_q} = Q(q,x^{(0)})$ with the decision-relevant questions that should be answered.

    \item The \textbf{answer agent} $A:X\times\mathcal{Q}^* \rightarrow \mathcal{A}^*$ looks at the company features $x^{(d)}$ and uses them to generate answers $\{a_l\}_{l=1}^{M_q}$ to all questions in the tree. It also has access to a web search tool that it can use agentically. Like \textit{ProbTree} \citep{cao2023probtree}, it answers question trees in a post-order traversal, aggregating answers from child nodes when generating answers for parent nodes. We provide further details in Appendix \ref{sec:appendix_question_trees}.
\end{itemize}

When executed for all seed questions, the agents produce a rich hierarchically structured \textbf{fact base} $F \subset \mathcal{F}$ ($\mathcal{F}$ is the universe of all possible question-answer pairs) about the company $x$:

$$F = \bigcup\limits_{q \in Q_0} \{(q_l, a_l)\}_{l=1}^{M_q}$$

\subsection{Reasoning Phase}
In the reasoning phase, DIALECTIC combines facts (possibly from different question trees) into arguments taking a stance on whether the VC should invest in a company or not. Let $s\in \{+,-\}$ denote the pro or contra stance, $\mathcal{G}$ the universe of natural-language arguments, and $\mathcal{C}$ the universe of natural-language critiques of these arguments. We define four LLM-based agent operations:

\begin{itemize}
    \item The \textbf{generator} $G^{(s)}:\mathcal{F}\rightarrow \mathcal{G}^K$ takes the fact base and generates $K$ \textbf{arguments} $\{g^{(s)}_j\}_{j=1}^K=G^{(s)}(F)$ per stance $s$, citing various facts from the fact base. This is inspired by \citet{park2023generativeagents} who recursively synthesize observations into higher-level reflections.

    \item The \textbf{critic} $C^{(s)}:\mathcal{F} \times \mathcal{G} \rightarrow \mathcal{C}$ criticizes an argument, producing a \textbf{critique} $c_j^{(s)}=C^{(s)}(F,g_j^{(s)})$ against it, possibly also citing facts from the fact base. The critic thereby acts as a \textit{devil's advocate} \citep{kim2024debate} sparking a debate about the argument.

    \item The \textbf{evaluator} $E:\mathcal{G} \times \mathcal{C} \rightarrow \mathbb{N}$ takes an argument and corresponding critique and judges the convincingness of the argument with a \textbf{quality score} $\sigma^{(s)}_j=E(g^{(s)}_j,c^{(s)}_j) \in \mathbb{N}$. Internally, it uses a 14-criteria evaluation scheme based on the argument quality taxonomy by \citet{wachsmuth2017argumentquality}. See Appendix \ref{sec:appendix_argument_evaluation} for further details.

    \item The \textbf{refiner} $R^{(s)}: \mathcal{F} \times \mathcal{G} \times \mathbb{N} \rightarrow \mathcal{G}$ refines a given argument trying to improve its quality. It produces a \textbf{refined argument} $\tilde{g}^{(s)}_j=R^{(s)}(F,g^{(s)}_j,\sigma^{(s)}_j)$.
\end{itemize}

As the refinement can be repeated, we use the notation $g^{(s)}_{j,t+1}=R^{(s)}(F,g^{(s)}_{j,t},\sigma^{(s)}_{j,t})$ instead, where the index $t=1,...,T$ denotes the iteration.

Starting with an initial set of $K_0$ arguments $\{g^{(s)}_{j,0}\}_{j=1}^{K_0}=G^{(s)}(F)$, DIALECTIC iteratively critiques, evaluates, and refines the arguments. It hereby follows a \textit{survival-of-the-fittest} logic, keeping only the best $K_t$ arguments (the \textbf{survivors} $S_t$) after each iteration $t$:

$$S^{(s)}_{t+1}=TopK(\{g^{(s)}_{j,t+1}: g^{(s)}_{j,t} \in S^{(s)}_{t}\},K_{t+1}),$$

where $TopK(\{\cdot\},K_{t+1})$ denotes the $K_{t+1}$ arguments with the highest quality scores $\sigma^{(s)}_{j,t+1}$ in $\{\cdot\}$. With arguments iteratively improving and $K_t$ decreasing over the iterations, DIALECTIC converges to a narrow selection $S_T=S^{(+)}_T \cup S^{(-)}_T$ of high-quality pro and contra arguments, where $|S_T^{(+)}| = |S_T^{(-)}| = K_T$. 
This mimics a debate in a VC investment committee where different members have different stances on the investment and continue to bring forward arguments and critiques of other members' arguments until the room converges to a dominant narrative.

\subsection{Decision-Making Phase}
After $T$ iterations of debate, a few dominant arguments for either stance have emerged. To determine which stance has the better arguments, we look at the sum of the argument quality scores for all surviving arguments and compare the pro and contra stances, calculating the \textbf{decision score} $\sigma_T$:

$$\sigma_T=\sigma^{(+)}_T - \sigma^{(-)}_T + \tau,$$

where $\sigma^{(s)}_T = \sum \sigma^{(s)}_{j,T}$ is the sum of the quality scores of all surviving arguments ${g^{(s)}_{j,T} \in S^{(s)}_T}$ and $\tau$ is a \textbf{decision threshold} capturing VC's preference for a margin of safety. Finally, DIALECTIC will decide to invest if and only if $\sigma_T > 0$.

\subsection{Hyperparameters \& Implementation}
The above definition of DIALECTIC presents three main hyperparameters: The number of arguments kept per iteration $K_t$, the number of iterations $T$, and the decision threshold $\tau$. In our implementation we set $K_t=5$ for $t \ne T$ and test different values of $K_T$, $T$, and $\tau$. For the LLM, we use OpenAI's \verb|gpt-5-mini-2025-08-07| \citep{openai2025gpt5mini}. We set the \verb|temperature| parameter to 0.0 for the answer agent and to 0.5 for all other agents. We report all used prompts in Appendix \ref{sec:appendix_prompts}.

\section{Evaluation Setup}
\label{sec:evaluation_setup}
We evaluate our method in a backtesting experiment by predicting startup success from historic data and benchmarking against real VC investors. Our dataset includes 259 startups that were added to real VCs' watchlists\footnote{The watchlists comprises data of five different VC funds and the real VCs' performance reported in this paper represents a weighted average across these funds.} between January 1, 2021 and December 31, 2021. The VCs considered joining the initial funding rounds (\textit{seed} or \textit{pre-seed}
) of these startups, which were raised some time between January 1, 2021 and February 28, 2023.

\paragraph{Dependent variable}
Similar to prior work \citep{sharchilev2018startupsuccess, gavrilenko2023improvingstartupsuccess}, we define a startup as \textit{successful}, if it has subsequently raised a \textit{series A} or later round by September 1, 2025, otherwise as \textit{unsuccessful}. With startup success as the dependent variable, our setup is a binary classification. Among all 259 startups, 25\% were successful.
\begin{figure*} [ht!]
    \centering
    \includegraphics[width=1\linewidth]{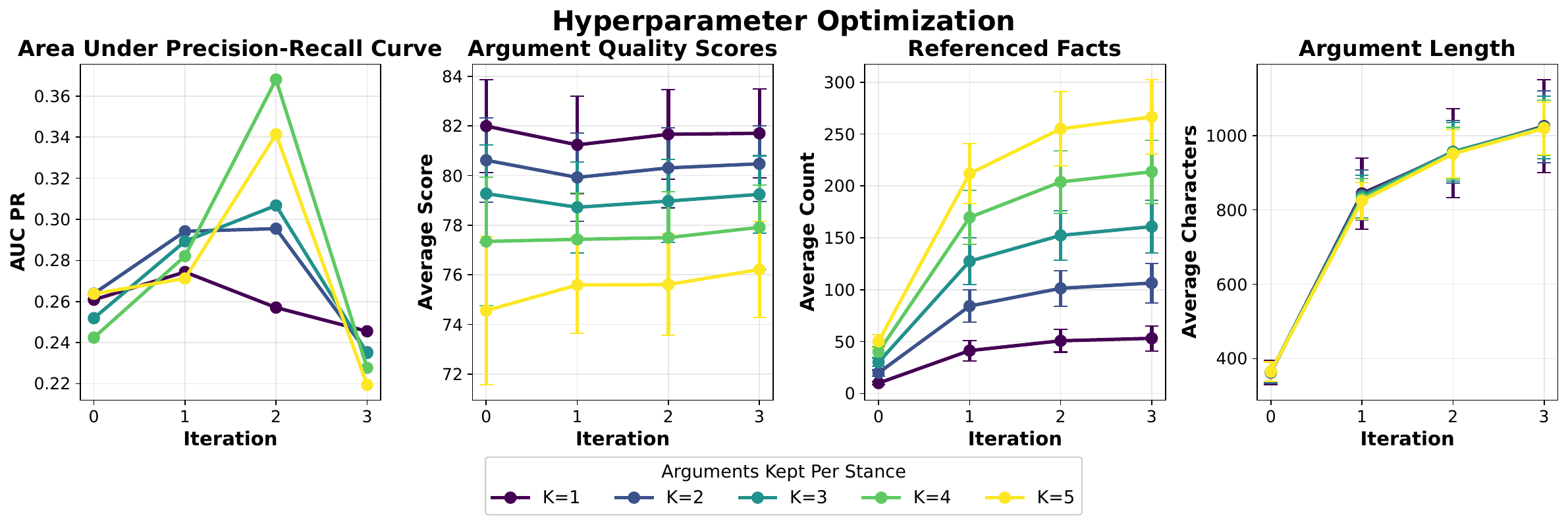}
    \caption{Results from the hyperparameter optimization, showing AUC-PR, raw argument scores, QA pair count, and argument length for different numbers of arguments $K_t$ and iterations $T$.}
    \label{fig:hyperparameter_optimization}
\end{figure*}
\paragraph{Independent variables}
To predict startup success, the following features are known for each startup: company name, short and long description, industry domain, team description, website content, and web search results (Table \ref{tab:data_fields} in the Appendix presents descriptions of all features). These features are extracted from the VCs' watchlists, \textit{Crunchbase.com}, startup homepages, and the \textit{Perplexity Sonar API}. To prevent \textit{look-ahead bias} \citep{zbikowski2021biasfree}, we use historic data snapshots and time filters to ensure all features have been available to the VC at the time of the investment decision. See Appendix \ref{sec:appendix_dataset} for a detailed description of the dataset and its creation.

\paragraph{Baselines}
We compare the classification performance of our method against the performance of the real VCs. The VCs invested in 6 of the 259 startups, and 2 of these were successful. Also, we compare against simple input-output (IO) prompting (see Listing \ref{lst:prompt_io_prompting} in the Appendix for the prompt).

\paragraph{Dataset split}
We split the data into a validation set with 129 startups and a test set with 130 startups using random stratified sampling, so that both sets have an equal ratio of successful startups and each set includes 3 startups that the VCs invested in.

\paragraph{Metrics}
To measure the performance of DIALECTIC and its baselines, we primarily look at precision and recall for different values of the decision threshold $\tau$, as well as the area under curve (AUC) of the precision-recall (PR) line. We also assess the argument quality scores, number of cited facts, length of the arguments, as well as the distribution of decision scores. We first identify a well-performing combination of our hyperparameters $T$ and $K_T$ on the validation set based on AUC-PR. We then evaluate this configuration on the test set.

\section{Results}
\label{sec:results}

Our results cover hyperparameter  optimization, comparative predictive performance, and an analysis of which facts DIALECTIC uses.

\subsection{Hyperparameter Optimization}

We optimize the system by varying the number of surviving arguments per side ($K_T$) and the number of refinement iterations ($T$). These parameters control how broadly the system explores arguments and how deeply it refines them. Figure \ref{fig:hyperparameter_optimization} summarizes the effect of varying these hyperparameters. Precision--recall performance shows a clear pattern: AUC-PR increases consistently from $T=0$ to $T=2$ and declines for $T \geq 3$. The best result occurs at $T=2$ with $K_2=4$, which we use for subsequent experiments.

Figure \ref{fig:hyperparameter_optimization} also displays coherent trends across other measures. Argument quality scores increase with more iterations, with only mild variation across $K_t$. Referenced facts rise with both parameters, suggesting stronger arguments rely on broader evidence. Argument length jumps from $T=0$ to $T=1$, driven by the introduction of structured justification, and grows more slowly thereafter as later iterations add elaboration rather than new insights. Because length and referenced facts increase smoothly while AUC-PR declines only at $T \geq 3$, the performance drop likely reflects over-refinement effects (e.g., redundancy or drift) rather than simple argument inflation.

\subsection{Predictive Performance Against Baselines}

\begin{figure} [ht!] \centering \includegraphics[width=1\linewidth]{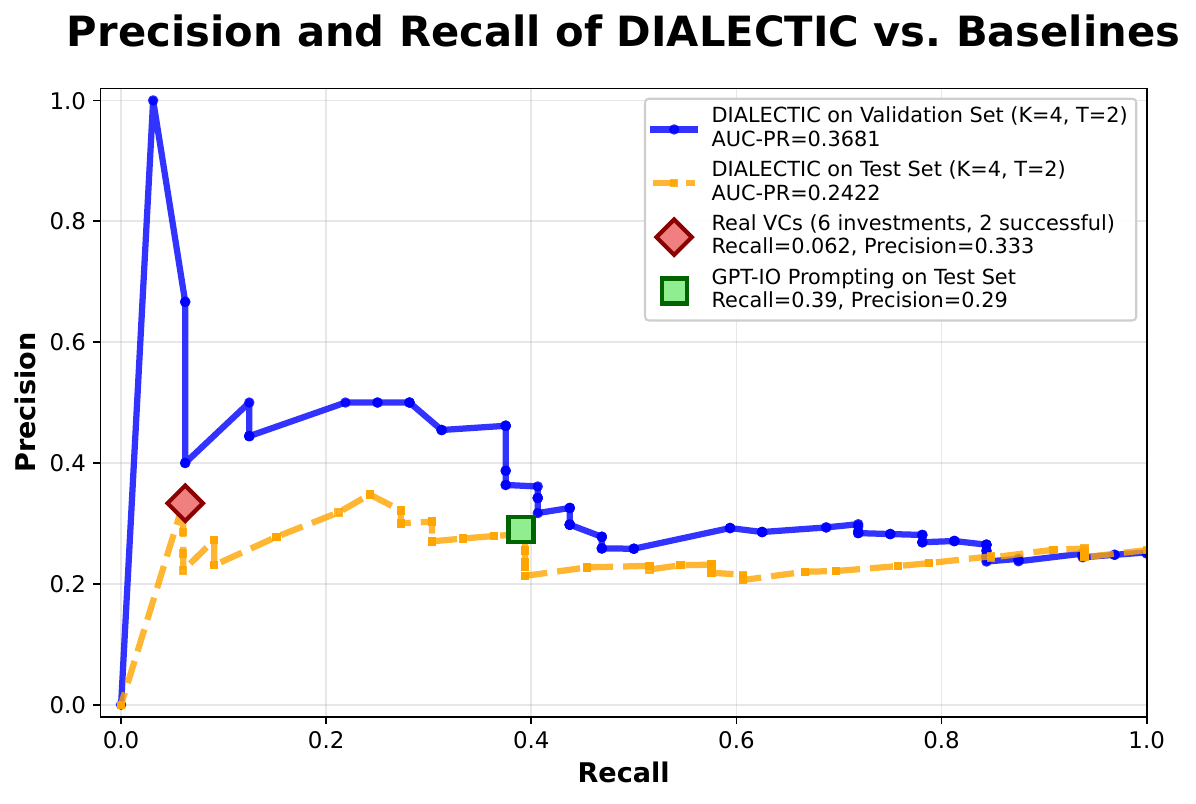} \caption{Precision and recall of DIALECTIC across all possible values of the decision threshold $\tau$ in comparison to the human VCs and GPT IO prompting baseline.} \label{fig:pr} \end{figure}

Figure~\ref{fig:pr} reports the predictive performance of DIALECTIC on the validation set and the held-out test set. The system attains an AUC-PR of 0.2422 on the test set, with precision comparable to human investors and the GPT-IO prompting baseline. Performance is higher on the validation set, where it achieved higher AUC-PR and even outperformed the real VCs. The operating points in Figure~\ref{fig:pr} show that, in the high-precision, low-recall region, the system behaves similarly to the baselines. Unlike the baselines, it produces a full ranked frontier rather than a single operating point, which allows practitioners to choose a decision threshold tailored to screening capacity. Overall, DIALECTIC is comparable to baselines in predictive performance while offering a full decision frontier and ranking rather than a single operating point.

\begin{figure} [ht!]
    \centering
    \includegraphics[width=1\linewidth]{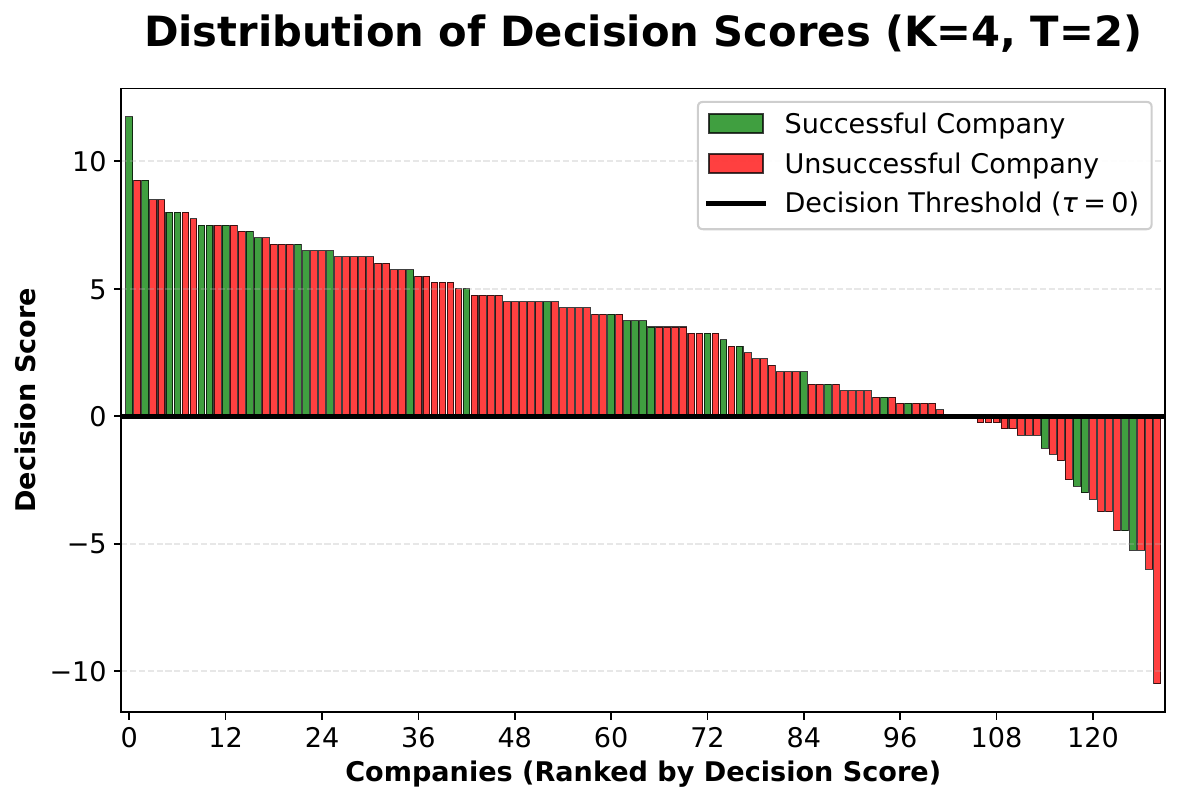}
    \caption{Distribution of decision scores.}
    \label{fig:decision_scores}
\end{figure}
Figure~\ref{fig:decision_scores} shows the distribution of decision scores for the best-performing configuration. Successful companies cluster at the top of the ranking, (left end of the plot), while unsuccessful ones appear toward the bottom (right side of the plot). This separation shows that higher scores correspond to a higher likelihood of success. In practice, selecting a threshold simply involves choosing a cutoff along this ranking. A higher threshold prioritizes the strongest opportunities and filters out most low-scoring cases.

\subsection{Evidence Utilization}

\begin{table}[ht!]
    \centering
    \setlength{\tabcolsep}{8pt} 
    \renewcommand{\arraystretch}{1.1} 
    \begin{tabular}{lccc}
        \toprule
        \textbf{Aspect} & \textbf{Usage} & \textbf{Availability} & \textbf{Ratio} \\
        \midrule
        General & 34.40\% & 34.88\% & 0.986 \\
        Team             & 20.52\% & 20.17\% & 1.017 \\
        Product          & 29.77\% & 29.43\% & 1.011 \\
        Market           & 15.31\% & 15.51\% & 0.987 \\
        \bottomrule
    \end{tabular}
    \caption{Utilization of factual evidence in arguments. Aspect refers to the question trees built for the four seed questions. Usage measures the share of all cited facts, availability measures the relative size of question trees, and ratio is the ratio of usage and availability.}
    \label{tab:aspect_usage}
\end{table}

Table~\ref{tab:aspect_usage} summarizes how the model uses different evidence categories when generating factual references. General company and product information dominate, accounting for nearly 65\% of all references, mirroring their share in the fact base. The ``ratio'' column reports the ratio between how often an aspect is referenced and its relative representation in the fact base. Values slightly above one (team: 1.017; product: 1.011) indicate that these aspects are referenced more frequently than their availability alone would predict. Market information is slightly under-used (ratio < 1). Overall, ratios cluster near one, indicating proportional use of available evidence. Notably, team-related evidence receives the highest relative usage, which aligns with established findings that investors frequently prioritize founder and team attributes when forming investment judgments \cite{gompers2020howvcsmake}.

\section{Conclusion}
\label{sec:conclusion}
This paper introduced DIALECTIC, an LLM-based multi-agent system for early-stage startup screening. The system integrates fact extraction, argument generation, iterative critique, and scoring into one pipeline. Evaluated on an industry-sourced dataset, the system achieves predictive performance comparable to historical human investment decisions while producing interpretable argument structures.

A central contribution of the approach is the introduction of iterative argumentation at the beginning of the investment funnel. Since investor bandwidth limits such deliberation during initial screening, it usually occurs later in the funnel. Enabling it earlier provides a structured foundation for preliminary assessments and supports reasoning under uncertainty.

Operationally, the system reduces time to initial assessment and produces a ranking when deal volume exceeds human screening capacity. It supports both fixed-threshold (returning only companies exceeding a certain decision score) and fixed-quantity (returning only the top N companies according to their decision scores) screening modes, reflecting constraints encountered in practice. Since missing strong opportunities is costlier than evaluating weak ones, the ranking mechanism aligns with recall-oriented objectives common in top-of-funnel screening. The generated arguments and evidence also support later stages such as due diligence or memo preparation.


\section*{Limitations}
\label{sec:limitations}
Venture capital is a domain of high uncertainty and low signal-to-noise ratios. This is reflected in the sensitivity of performance to hyperparameter choices and sample composition, highlighted by the difference in AUC-PR seen between the validation and the test sets.

In investment decision tasks, avoiding look-ahead bias is a critical priority. To minimize this and ensure feature completeness, we applied strict temporal and availability constraints during dataset construction. As a consequence, more than 90\% of the original companies were excluded, leaving 259 companies in the final dataset (see Appendix \ref{sec:appendix_dataset}). Similarly, this also restricted the corresponding baseline of real VCs' performance to data of only 6 invested companies.

While extensive measures were taken to minimize look-ahead bias, a residual risk remains due to the use of a single historic \textit{Crunchbase} snapshot from January 24, 2022. For companies announcing seed or pre-seed rounds prior to this date, some data fields may reflect information not strictly available at decision time. However, we consider it unlikely that this information provides substantial information about the future as the potentially affected data fields are limited to mostly static company data and do not provide any information on future funding rounds. 

At the same time, our GPT-IO prompting baseline may be affected by look-ahead bias, if \verb|GPT-5-mini| was trained on data about the companies in our dataset, possibly overstating the performance of this baseline. Long-term studies capturing reliable data and evaluating startup performance over a long timeframe would be needed to certainly rule out any risk of look-ahead bias.

As done by several previous studies (see Table \ref{tab:related_work_overview} in the Appendix for an overview), we modeled startup evaluation as a success prediction task with a binary definition of success. This abstraction does not entirely capture the multi-dimensional and time-dependent nature of real venture outcomes. Moreover, the business impact of a VC investor on a startup’s success is an unobservable counterfactual and cannot be evaluated in a retrospective backtesting study.

Lastly, the scope of our analysis is limited to early-stage VC investments (\textit{pre-seed}/\textit{seed} to \textit{series A}) in European companies and may not generalize to other forms of VC or private equity.

\section*{Future Research}
\label{sec:future research}

Several directions emerge for extending this work beyond the current evaluation setting. Evaluating argument-based screening systems across multiple historical data snapshots and longer time horizons would better separate decision-time information from realized outcomes, further reducing residual look-ahead risk and enabling analysis of how arguments evolve as evidence accumulates. 

Also, moving beyond binary success labels toward multi-level, time-dependent, or continuous outcome measures would better capture heterogeneous venture trajectories and support finer-grained assessment of decision quality under uncertainty. 

Furthermore, deeper analysis of debate dynamics, including agent role diversity, critique depth, turn structure, and stopping criteria, could shift evaluation from aggregate performance toward understanding when and why multi-agent argumentation improves reasoning or fails. 

Lastly, extending evaluation to later-stage investments, non-European markets, and adjacent decision contexts with noisy, unstructured evidence would help assess robustness and domain transferability beyond early-stage VC screening.

\section*{Acknowledgments}
This work was supported by research funding from Earlybird Venture Capital and UVC Partners. The authors further thank Earlybird Venture Capital for providing access to the historical investment data used in this analysis.



\appendix

\section{Dataset Creation}
\label{sec:appendix_dataset}
For our backtesting experiment, we prepared a dataset of real startups from the watchlists of five VC funds (in the following we will refer to these watchlists in singular). The dataset was created by extracting and merging data from four sources: (1) the \textbf{watchlist} of a real VC, (2) \textbf{Crunchbase} (\url{crunchbase.com}), an online database aggregating information about businesses, (3) historical snapshots of the \textbf{startup websites} retrieved through the \textit{Internet Archive}'s \textit{Wayback Machine} (\url{web.archive.org}), and (4) \textbf{search results} obtained through the Perplexity Sonar API. Table \ref{tab:data_fields} includes a summary of all extracted data fields and their origination dates.

\subsection{Preventing Look-Ahead Bias}
\label{sec:appendix_preventing_lookahead}
When working with historic data, it is important to consider which information was and was not available to the VC at the time the investment decision had to be made. Including information about the startups originating from a time after the VC's decision would constitute a \textit{look-ahead bias} \citep{zbikowski2021biasfree}. Doing so could put the backtested method at an unfair advantage compared to the real VC, because it could leak information about the startup's future success. To prevent such a bias, we carefully filter the information available during backtesting by using historic data snapshots.

\paragraph{Cutoff date}
In order for the real VC to participate in the initial funding round, the investment decision had to be made at some point between the VC becoming aware of the startup (the date the startup was added to the VC's watchlist) and the announcement of the initial funding round, likely closer to the latter. Therefore, we consider the announcement date of the initial funding round as a cutoff and use it to restrict the information that we include in the dataset:

\begin{itemize}
    \item \textbf{Watchlist:} We limit the data to startups that were added to the watchlist between January 1, 2021 and December 31, 2021 and had not yet raised a \textit{series A} or later by the time they were added.
    
    \item \textbf{Crunchbase:} We had access to two \textit{Crunchbase} snapshots taken on January 24, 2022 and September 1, 2025. We refer to these as the \textit{historic} and the \textit{current} snapshot, respectively. We use only the historic snapshot to extract predictive features. The current snapshot is used to determine whether a startup turned out successful (i.e., received subsequent funding).
    
    \item \textbf{Startup websites:} For each startup, we retrieve a historic snapshot of its website through \textit{Wayback Machine} from the latest available date before the announcement of the initial funding round.

    \item \textbf{Search results:} For the \textit{Perplexity Sonar API}, we apply a time filter to return only those search results that originate from a time before the announcement of the initial funding round.
\end{itemize}

\begin{table*}[htbp]
    \centering
    \begin{tabularx}{\textwidth}{@{} P{2.5cm} P{3.5cm} Y P{3cm} @{}}
        \toprule
        \textbf{Source} & \textbf{Data Field} & \textbf{Description} & \textbf{Value As Of} \\
        \midrule
        
        \multirow{1}{*}{Watchlist} 
          & \textit{Name} \fsymbol{}
          & The company name.
          & Some time in 2021 \\
    
        \addlinespace
        
          & \textit{Domain} \fsymbol{}
          & The company web domain.
          & Some time in 2021 \\
    
        \addlinespace
        
          & \textit{Date Added} \psymbol{}
          & The date the company was added to the watchlist. Only companies added between January 1, 2021 and December 31, 2021 are considered.
          & Some time in 2021 \\
    
        \addlinespace
        
          & \textit{Status} \psymbol{}
          & The last stage reached in the investment process (e.g., \textit{Added to Watchlist}, \textit{Initial Review}, or \textit{Investment Made}). Used to determine whether the real VC invested in the company. 
          & Some time in 2021 \\
        
        \midrule
        
        \multirow{1}{*}{Crunchbase}
          & \textit{Funding Rounds} \psymbol{}
          & A list of all funding rounds, including round type (\textit{pre-seed}, \textit{seed}, \textit{series A}, \textit{IPO}, etc.), amount, and announcement date of the round.
          & September 1, 2025 \\
    
        \addlinespace
        
          & \textit{Current Name} \psymbol{}
          & The company’s current name. 
          & September 1, 2025 \\
    
        \addlinespace
        
          & \textit{Current Domain} \psymbol{}
          & The company’s current web domain.
          & September 1, 2025 \\
        
        \midrule
        
        \multirow{1}{*}{Crunchbase}
          & \textit{Short Description} \fsymbol{}
          & Short description of the company. 
          & January 24, 2022 \\
    
        \addlinespace
        
          & \textit{Long Description} \fsymbol{}
          & Long description of the company. 
          & January 24, 2022 \\
    
        \addlinespace
        
          & \textit{Industries} \fsymbol{}
          & A list of industries the company is operating in.
          & January 24, 2022 \\
    
        \addlinespace
        
          & \textit{Team} \fsymbol{}
          & The names of the team members, their education, prior work experience, and current roles. 
          & January 24, 2022 \\
    
        \addlinespace
        
          & \textit{Historic Name} \psymbol{}
          & The company’s former name. 
          & January 24, 2022 \\
    
        \addlinespace
        
          & \textit{Historic Domain} \psymbol{}
          & The company’s former web domain. 
          & January 24, 2022 \\
    
        \addlinespace
        
          & \textit{Historic Funding Rounds} \psymbol{}
          & A list of all historic funding rounds, including round type, amount, and announcement date.
          & January 24, 2022 \\
        
        \midrule
        
        \multirow{1}{*}{Startup Websites} 
          & \textit{Website} \fsymbol{}
          & The historic HTML content of the company’s website (only the homepage).
          & Various dates \\
    
        \addlinespace
        
          & \textit{Archived} \psymbol{}
          & The date and time when the website was captured. 
          & Various dates \\
        
        \midrule
        
        \multirow{1}{*}{Search Results} 
          & \textit{Results} \fsymbol{}
          & The list of search results for a given query, including title, content snippet, and URL.
          & Various dates \\
        
        \bottomrule
    \end{tabularx}
    \caption{An overview of the data sources and extracted data fields used for the dataset creation. Data fields marked with F are used as predictive features (independent variables), whereas data fields marked with P are only used during preprocessing (e.g., for merging data from different sources) and were \textbf{not} made available to the prediction method.}
    \label{tab:data_fields}
\end{table*}

\subsection{Data Preprocessing}
To construct the dataset, we started with companies that were added to the VC's watchlist between January 1, 2021 and January 31, 2021 and systematically enriched this set with data from \textit{Crunchbase}, startup websites, and search results, removing companies where such enrichments were not possible. The preprocessing followed four phases: (1) data cleaning, (2) entity matching, (3) label assignment, and (4) enrichment. Table \ref{tab:preprocessing} reports the number of companies retained after each step and the corresponding share of successful companies.

\begin{enumerate}
    \item \textbf{Watchlist export:} The starting dataset contained 3,441 companies from the VC's watchlist that were between January 1, 2021 and January 31, 2021.

    \item \textbf{Cleaning:} We cleaned the dataset by removing duplicate entries, \textit{Missed Deals}, and companies that were considered for a founding round later than \textit{seed}. We further updated all companies website URLs by sending HTTP requests to the domains listed in the watchlist export and recording the final redirect target as the current domain. Finally, we canonicalized company names and URLs (lowercasing, removing prefixes such as ``www'', unicode normalization) for later matching purposes. After cleaning, the dataset contained 3,357 companies.

    \item \textbf{Entity matching:} To enrich the watchlist records with additional data from \textit{Crunchbase}, we performed a left-join between the cleaned dataset and the current \textit{Crunchbase} snapshot and then the historic \textit{Crunchbase} snapshot. Matching followed a strict precedence: (1) exact current domain match, (2) exact historic domain match, and (3) fuzzy name-and-domain match with a similarity threshold of 95\%. After matching, the dataset contained 1,623 companies.

    \item \textbf{Label assignment:} Success labels were constructed from the current \textit{Crunchbase} snapshot. Companies that had raised a series A or later funding round by September 1, 2025 were labeled as \textit{successful}, all others as \textit{unsuccessful}.

    \item \textbf{Enrichment:} We further enriched each startup's data with additional historic information from the startup's website, web search results, and historic \textit{Crunchbase} snapshot to extract predictive features including long and short company descriptions, industry, team setup, website content, and information from online articles. We removed companies where no founding team information was available, leaving 637 companies.
\end{enumerate}

\begin{table*}[htbp]
    \centering
    \begin{tabular}{@{} l rr @{}}
        \toprule
        \textbf{Preprocessing Stage} & \textbf{Companies} & \textbf{Success Rate (\%)} \\
        \midrule
        Export from CRM system                  & 3,441 & --- \\
        Remove duplicates and missed deals                & 3,404 & --- \\
        Remove series A investments by VC funds                   & 3,401 & --- \\
        Match with current funding data                   & 2,192 & 21.6 \\
        Remove companies without cutoff date              & 1,715 & 23.4 \\
        Remove companies added post–seed announcement     &   587 & 18.7 \\
        Apply temporal cutoff (before February 28, 2023)         &   462 & 22.0 \\
        Match with historic Crunchbase snapshot           &   259 & 25.1 \\
        \bottomrule
    \end{tabular}
    \caption{Dataset size and success rate after successive preprocessing steps. The success rate refers to the share labeled \textit{successful} at the corresponding stage.}
    \label{tab:preprocessing}
\end{table*}

\section{DIALECTIC Design Details}

\subsection{Seed Questions}
\label{sec:appendix_seed_questions}
To kick off DIALECTIC's question decomposition, we provide it with a set of seed questions $Q_0$. These are intended to guide DIALECTIC's fact gathering efforts by giving it a rough scaffolding of relevant fact categories. VC investors typically assess startups across the following dimensions: general company, market, product/service, entrepreneurial team, and funding \citep{retterath2020humanvscomp}. While rich and reliable funding information is typically private and was not available to us for every startup in our dataset, we dedicate one seed question to each of the remaining four dimensions. Specifically, we use the following four seed questions:

\begin{enumerate}
    \item \textbf{General Company:} ``How do the company’s sector, development stage, and geography align with the VC’s investment strategy?''

    \item \textbf{Team:} ``Who are the key members of the founding team, and what relevant experience and track record do they have?''

    \item \textbf{Product:} ``What are the product’s core features, underlying technology, and forms of protection?''
    
    \item \textbf{Market:} ``What is the current size, historical growth, and forecast growth of the target market, and which customer needs or market gaps does the company address?''
\end{enumerate}

\subsection{Generating Question Trees}
\label{sec:appendix_question_trees}

In order to evaluate each startup in more detail, DIALECTIC decomposes each seed question into lower-level questions tailored to the specific industry of the startup. Together, the seed questions and all their lower-level questions are supposed to comprehensively cover the information required by the VC investor to make a decision. In order to derive and answer lower-level questions from the seed questions, we adapt the \textit{Probabilistic Tree-of-Thought Prompting} (\textit{ProbTree}) approach \citep{cao2023probtree}. ProbTree uses an LLM to create hierarchical question decomposition trees (HQDTs) and then answer the questions in a post-order traversal using three different answer strategies \textbf{Open Book} (retrieving information from online sources), \textbf{Closed Book} (asking an LLM for its internal knowledge), and \textbf{Child Aggregation} (deriving the answer to a higher-level question from the answers to its lower level questions). For each answer strategy, it calculates a confidence score and then probabilistically chooses the most confident answers for each question.

For DIALECTIC, we use a simplified adaptation of ProbTree. It first decomposes a given seed question into a HQDT in a single LLM prompt. Then it performs a post-order traversal through the tree to answer the questions from leaf nodes to the root node. Unlike ProbTree, we use a single answer prompt for each node and therefore forgo the confidence estimation. Each prompt includes a company summary (description, tagline, and team details including education and prior work experience) and optional web data that the LLM can obtain agentically, if it decides to do so, by using a web search tool. We only allow usage of the web search tool for leaf nodes.

For the web searches, we provide the LLM with access to the \textit{Perplexity Sonar API}. We limit search results to five, each described by a title and content snippet. As described in Appendix \ref{sec:appendix_preventing_lookahead}, we restrict search results to those originating from a time before the announcement date of the initial funding round to prevent look-ahead bias.

\subsection{Evaluating Arguments}
\label{sec:appendix_argument_evaluation}
DIALECTIC includes an evaluator agent that assigns a numeric quality score for each argument. We use an LLM judge \citep{zheng2023llmasajudge} to evaluate each argument and apply the taxonomy of argument quality proposed by \citet{wachsmuth2017argumentquality}. Following the instruction design principles by \citet{wachsmuth2024argumentllm}, we adapt the taxonomy criteria to the VC context. The revised framework explicitly defines the objective of argumentation (informing the investment decision), establishes domain-specific criteria for argument quality, specifies the intended audience (expert VC investors), and incorporates the surrounding decision context (high-stakes financial environments). Our argument quality evaluation scheme includes the following 14 questions:

\begin{enumerate}
    \item \textbf{Local Acceptability:} Are the premises believable and factually plausible given the provided Q\&A facts?
    \item \textbf{Local Relevance:} Do the premises clearly contribute to supporting or rejecting the conclusion about investment?
    \item \textbf{Local Sufficiency:} Do the premises provide enough support to justify the conclusion?
    \item \textbf{Cogency:} Does the argument have premises that are acceptable, relevant, and sufficient to support the investment conclusion?
    \item \textbf{Credibility:} Does the argument make the author appear credible and trustworthy to VC investors?
    \item \textbf{Emotional Appeal:} Does the argument create emotions that make the VC investors more receptive?
    \item \textbf{Clarity:} Does the argument use correct and widely unambiguous language as well as avoid deviation from the issue?
    \item \textbf{Appropriateness:} Is the style of reasoning and language suitable for a professional VC investment discussion?
    \item \textbf{Arrangement:} Is the argument well-structured, with a logical order of premises and conclusion?
    \item \textbf{Effectiveness:} Does the argument succeed in persuading the VC investors toward or against investing?
    \item \textbf{Global Acceptability:} Would most VCs consider it a valid and legitimate argument?
    \item \textbf{Global Relevance:} Does the argument meaningfully contribute to resolving the overall investment question?
    \item \textbf{Global Sufficiency:} Does the argument adequately anticipate and rebut the main counterarguments from the argument’s stance?
    \item \textbf{Reasonableness:} Does the argument resolve the issue in a way acceptable to the VC investors, balancing global acceptability, relevance, and sufficiency?
\end{enumerate}

The LLM judge scores each argument across the above 14 criteria using a seven-point Likert scale from 1 (low) to 7 (high). To calculate the final argument quality score, we simply sum up all of the 14 individual scores. The judge also produces justifications explaining each score.

\section{Prompts}
\label{sec:appendix_prompts}
In the following, we report the prompts that we used for each of our LLM-based agents. These are:

\begin{itemize}
    \item \textbf{Decomposer} prompt (Listing \ref{lst:prompt_decomposer})
    \item \textbf{Answer Agent} prompt (Listing \ref{lst:prompt_answer})
    \item \textbf{Generator} prompt for pro (Listing \ref{lst:prompt_generator_pro}) and for contra arguments (Listing \ref{lst:prompt_generator_contra})
    \item \textbf{Critic} prompt for pro (Listing \ref{lst:prompt_critic_pro}) and for contra arguments (Listing \ref{lst:prompt_critic_contra})
    \item \textbf{Evaluator} prompt (Listing \ref{lst:prompt_evaluator})
    \item \textbf{Refiner} prompt for pro (Listing \ref{lst:prompt_refiner_pro}) and for contra arguments (Listing \ref{lst:prompt_refiner_contra})
    \item \textbf{Input Output (IO) Prompting} baseline prompt (Listing \ref{lst:prompt_io_prompting})
\end{itemize}

\begin{lstlisting}[
  basicstyle=\ttfamily\small,
  frame=single,
  breaklines=true,
  breakatwhitespace=false,
  columns=fullflexible,
  keepspaces=true,
  caption={Decomposer Prompt},
  label={lst:prompt_decomposer}
]
SYSTEM: You are good at decomposing a complex question into a hierarchical question decomposition tree (HQDT).

USER: Please generate a hierarchical question decomposition tree (HQDT) with json format for a given question. In this tree, the root node is the original complex question, and each non-root node is a sub-question of its parent.

Q: How large is the company's market opportunity (TAM, SAM, SOM)?
A: {{
  "How large is the company's market opportunity (TAM, SAM, SOM)?": [
    "What is the Total Addressable Market (TAM)?",
    "What is the Serviceable Available Market (SAM)?",
    "What is the Serviceable Obtainable Market (SOM)?"
  ],
  "What is the Total Addressable Market (TAM)?": [
    "What customer segments are included in the broadest market?",
    "What is the total number of potential customers?",
    "What is the total industry revenue across those segments?"
  ],
  "What is the Serviceable Available Market (SAM)?": [
    "Which subset of TAM does the company's product or service directly target?",
    "What portion of customers can realistically be reached given geography, regulations, or product scope?",
    "What is the annual spending of these customers?"
  ],
  "What is the Serviceable Obtainable Market (SOM)?": [
    "What portion of SAM can the company realistically capture in the next 3-5 years?",
    "What customer acquisition assumptions support this share?",
    "What expected adoption rate drives this forecast?",
    "What annual revenue corresponds to this achievable market share?"
  ]
}}

Q: What is the competitive landscape, and how is the company positioned within it?
A: {{
  "What is the competitive landscape, and how is the company positioned within it?": [
    "What is the competitive landscape?",
    "How is the company positioned within the competitive landscape?"
  ],
  "What is the competitive landscape?": [
    "Who are the direct competitors?",
    "Who are the indirect competitors or substitutes?",
    "What are the major trends shaping competition in this space?"
  ],
  "How is the company positioned within the competitive landscape?": [
    "What is the company's relative pricing strategy?",
    "What is the company's market share or traction compared to peers?",
    "Does the company occupy a niche or broader category?",
    "What barriers to entry protect the company's position?"
  ]
}}

Q: What is the company's product differentiation and value proposition?
A: {{
  "What is the company's product differentiation and value proposition?": [
    "What is the company's product differentiation?",
    "What is the company's value proposition?"
  ],
  "What is the company's product differentiation?": [
    "What features or technologies distinguish the product?",
    "How is the product better than alternatives?",
    "What intellectual property (e.g., patents, proprietary tech) supports defensibility?"
  ],
  "What is the company's value proposition?": [
    "What problem does the product solve for customers?",
    "What measurable benefits (e.g., cost savings, time savings, revenue uplift) does it deliver?",
    "Why would customers choose this company over competitors?"
  ]
}}

Here is the question to decompose:
Q: {question}

Generate its HQDT customized for a company in the {industry} industry.
\end{lstlisting}

\begin{lstlisting}[
  basicstyle=\ttfamily\small,
  frame=single,
  breaklines=true,
  breakatwhitespace=false,
  columns=fullflexible,
  keepspaces=true,
  caption={Answer Agent Prompt},
  label={lst:prompt_answer}
]
SYSTEM: Answer the question using company summary and sub Q&A if provided. Keep answer concise (<50 words) with data backing.
If unable to answer the question, use web_search for market data, trends, competitive analysis, funding info. Focus on industry-level searches, not specific companies. Use the tool only if necessary.
Make ONE tool call at a time.

USER: Question: {question}

Company summary: {company_summary}
{qa_pairs}
\end{lstlisting}

\begin{lstlisting}[
  basicstyle=\ttfamily\small,
  frame=single,
  breaklines=true,
  breakatwhitespace=false,
  columns=fullflexible,
  keepspaces=true,
  caption={Generator Prompt (Pro Arguments)},
  label={lst:prompt_generator_pro}
]
SYSTEM: You are a very experienced investor at a top-tier VC fund. You are also a great storyteller and can tell a compelling story.

USER: Generate {n_pro_arguments} pro arguments why this company is a good investment opportunity.

Each argument should be concise (max. 100 words) and backed by specific data from the questions and answers.

A good argument provides a unique perspective on the investment opportunity that addresses the following criteria:
1. Local Acceptability - Are the premises believable and factually plausible given the provided Q&A facts?
2. Local Relevance - Do the premises clearly contribute to supporting or rejecting the conclusion about investment?
3. Local Sufficiency - Do the premises provide enough support to justify the conclusion?
4. Cogency - Does the argument have premises that are acceptable, relevant, and sufficient to support the investment conclusion?
5. Credibility - Does the argument make the author appear credible and trustworthy to VC investors?
6. Emotional Appeal - Does the argument create emotions that make the VC investors more receptive?
7. Clarity - Does the argument use correct and widely unambiguous language as well as avoid deviation from the issue?
8. Appropriateness - Is the style of reasoning and language suitable for a professional VC investment discussion?
9. Arrangement - Is the argument well-structured, with a logical order of premises and conclusion?
10. Effectiveness - Does the argument succeed in persuading the VC investors toward or against investing?
11. Global Acceptability - Would most VCs consider it a valid/legitimate argument?
12. Global Relevance - Does the argument meaningfully contribute to resolving the overall investment question?
13. Global Sufficiency - Does the argument adequately anticipate and rebut the main counterarguments from the argument's stance?
14. Reasonableness - Does the argument resolve the issue in a way acceptable to the VC investors, balancing global acceptability, relevance, and sufficiency?

Here are the questions and answers about the company:
{qa_pairs}

Provide the qa_indices that were used to generate the argument.
\end{lstlisting}

\begin{lstlisting}[
  basicstyle=\ttfamily\small,
  frame=single,
  breaklines=true,
  breakatwhitespace=false,
  columns=fullflexible,
  keepspaces=true,
  caption={Generator Prompt (Contra Arguments)},
  label={lst:prompt_generator_contra}
]
SYSTEM: You are a very experienced investor at a top-tier VC fund. You are also a great storyteller and can tell a compelling story.

USER: Generate {n_contra_arguments} contra arguments why this company is a bad investment opportunity.

Each argument should be concise (2-3 sentences) and backed by specific data from the questions and answers.
Lack of data is not a good contra argument.

A good argument provides a unique perspective on the investment opportunity that addresses the following criteria:
1. Local Acceptability - Are the premises believable and factually plausible given the provided Q&A facts?
2. Local Relevance - Do the premises clearly contribute to supporting or rejecting the conclusion about investment?
3. Local Sufficiency - Do the premises provide enough support to justify the conclusion?
4. Cogency - Does the argument have premises that are acceptable, relevant, and sufficient to support the investment conclusion?
5. Credibility - Does the argument make the author appear credible and trustworthy to VC investors?
6. Emotional Appeal - Does the argument create emotions that make the VC investors more receptive?
7. Clarity - Does the argument use correct and widely unambiguous language as well as avoid deviation from the issue?
8. Appropriateness - Is the style of reasoning and language suitable for a professional VC investment discussion?
9. Arrangement - Is the argument well-structured, with a logical order of premises and conclusion?
10. Effectiveness - Does the argument succeed in persuading the VC investors toward or against investing?
11. Global Acceptability - Would most VCs consider it a valid/legitimate argument?
12. Global Relevance - Does the argument meaningfully contribute to resolving the overall investment question?
13. Global Sufficiency - Does the argument adequately anticipate and rebut the main counterarguments from the argument's stance?
14. Reasonableness - Does the argument resolve the issue in a way acceptable to the VC investors, balancing global acceptability, relevance, and sufficiency?

Here are the questions and answers about the company:
{qa_pairs}

Provide the qa_indices that were used to generate the argument.
\end{lstlisting}

\begin{lstlisting}[
  basicstyle=\ttfamily\small,
  frame=single,
  breaklines=true,
  breakatwhitespace=false,
  columns=fullflexible,
  keepspaces=true,
  caption={Critic Prompt (Pro Arguments)},
  label={lst:prompt_critic_pro}
]
SYSTEM: You are a very experienced VC investor against investing in the company. However, your colleague thinks it is a good investment opportunity. 
Your job is to criticize the pro argument given by your colleague using the questions and answers about the company and defend your position.
Be direct to persuade your colleague not to invest in the company.

USER: Here are the questions and answers about the company:
{qa_pairs}

Here is the argument you have to criticize to persuade the colleague not to invest in the company:
{argument}

Keep your critique concise in 3-4 sentences.
\end{lstlisting}

\begin{lstlisting}[
  basicstyle=\ttfamily\small,
  frame=single,
  breaklines=true,
  breakatwhitespace=false,
  columns=fullflexible,
  keepspaces=true,
  caption={Critic Prompt (Contra Arguments)},
  label={lst:prompt_critic_contra}
]
SYSTEM:You are a very experienced VC investor in favor of investing in the company. However, your colleague thinks it is a bad investment opportunity. 
Your job is to criticize the given contra argument given by your colleague using the questions and answers about the company and defend your position.
Be direct to persuade your colleague to invest in the company.

USER: Here are the questions and answers about the company:
{qa_pairs}

Here is the argument you have to criticize to persuade the colleague to invest in the company:
{argument}

Keep your critique concise in 3-4 sentences.
\end{lstlisting}

\begin{lstlisting}[
  basicstyle=\ttfamily\small,
  frame=single,
  breaklines=true,
  breakatwhitespace=false,
  columns=fullflexible,
  keepspaces=true,
  caption={Evaluator Prompt},
  label={lst:prompt_evaluator}
]
SYSTEM: You are an impartial LLM judge to evaluate the quality of an argument in the VC investment context. The goal of the argument is to support or reject a startup investment decision in a persuasive way. 
The quality of an argument in the venture capital investment context should be evaluated along the following 14 dimensions. For each dimension, assign a score from 1 (Low) to 7 (High), and provide a short feedback (1 sentence) how to improve the score.

14 Dimensions:
1. Local Acceptability - Are the premises believable and factually plausible given the provided Q&A facts?
2. Local Relevance - Do the premises clearly contribute to supporting or rejecting the conclusion about investment?
3. Local Sufficiency - Do the premises provide enough support to justify the conclusion?
4. Cogency - Does the argument have premises that are acceptable, relevant, and sufficient to support the investment conclusion?
5. Credibility - Does the argument make the author appear credible and trustworthy to VC investors?
6. Emotional Appeal - Does the argument create emotions that make the VC investors more receptive?
7. Clarity - Does the argument use correct and widely unambiguous language as well as avoid deviation from the issue?
8. Appropriateness - Is the style of reasoning and language suitable for a professional VC investment discussion?
9. Arrangement - Is the argument well-structured, with a logical order of premises and conclusion?
10. Effectiveness - Does the argument succeed in persuading the VC investors toward or against investing?
11. Global Acceptability - Would most VCs consider it a valid/legitimate argument?
12. Global Relevance - Does the argument meaningfully contribute to resolving the overall investment question?
13. Global Sufficiency - Does the argument adequately anticipate and rebut the main counterarguments from the argument's stance?
14. Reasonableness - Does the argument resolve the issue in a way acceptable to the VC investors, balancing global acceptability, relevance, and sufficiency?

USER: Argument to evaluate:
{argument}
{critique}
...
\end{lstlisting}

\begin{lstlisting}[
  basicstyle=\ttfamily\small,
  frame=single,
  breaklines=true,
  breakatwhitespace=false,
  columns=fullflexible,
  keepspaces=true,
  caption={Refiner Prompt (Pro Arguments)},
  label={lst:prompt_refiner_pro}
]
SYSTEM: You are a very experienced investor at a top-tier VC fund. You are sure that the company is a good investment opportunity.
Your job is to revise your argument to reach better argument quality scores. 

USER: Here are the Q&A facts about the company:
{qa_pairs}

Here is your previous argument: 
{argument}

Here are the argument quality scores (1-7) to your previous argument: 
{argument_feedback}

Refine your argument by improving argument quality scores.
\end{lstlisting}

\begin{lstlisting}[
  basicstyle=\ttfamily\small,
  frame=single,
  breaklines=true,
  breakatwhitespace=false,
  columns=fullflexible,
  keepspaces=true,
  caption={Refiner Prompt (Contra Arguments)},
  label={lst:prompt_refiner_contra}
]
SYSTEM: You are a very experienced investor at a top-tier VC fund. You are sure that the company is a bad investment opportunity.
Your job is to revise your argument to reach better argument quality scores.

USER: Here are the Q&A facts about the company:
{qa_pairs}

Here is your previous argument: 
{argument}

Here are the argument quality scores (1-7) to your previous argument: 
{argument_feedback}

Refine your argument by improving argument quality scores.
\end{lstlisting}

\begin{lstlisting}[
  basicstyle=\ttfamily\small,
  frame=single,
  breaklines=true,
  breakatwhitespace=false,
  columns=fullflexible,
  keepspaces=true,
  caption={Input Output (IO) Prompting Baseline Prompt},
  label={lst:prompt_io_prompting}
]
SYSTEM: Assuming you are a venture capital investor, would you invest in the following company? Respond with only "Yes" or "No".

USER: Questions and Answers for the company:
{qa_pairs}
\end{lstlisting}

\section{Related Work}
\label{sec:appendix_related_work}
Table \ref{tab:related_work_overview} provides an overview of related work, i.e., studies that propose startup success prediction methods based on machine learning. The table also shows the success criteria used by these studies, the accuracy, recall, and precision achieved by the best-performing models, as well as the interpretability approach taken, if any.

\begin{table*}
    \centering
    \small
    \begin{tabular}{@{}p{3cm}p{3.2cm}p{1.5cm}p{1.5cm}p{1.5cm}p{3cm}@{}}
    \toprule
    \textbf{Work} & \textbf{Success Criterion} & \textbf{Accuracy} & \textbf{Recall} & \textbf{Precision} & \textbf{Interpretability} \\
    \midrule
    \citet{arroyo2019assessment} & First event in 3 yrs (AC = acquired, FR = funding round, IPO, CL = closed, NE = no event) & Global 82.2\% & FR: 40\%, AC: 3\%, IPO: very low, NE: 95\% & FR: 64\%, AC: 33\%, IPO: 44\%, NE: 85\% & Feature-based \\
    
    \addlinespace
    
    \citet{zbikowski2021biasfree} & AC, IPO, Series B & 85\% & 34\% & 57\% & Feature-based \\
    
    \addlinespace
    
    \citet{retterath2020humanvscomp} & Follow-on round, trade sale, IPO & 80\% & 80\% & -- & No mention \\
    
    \addlinespace
    
    \citet{antretter2019predicting} & 5-year survival & 76\% & 86\% & 80\% & Feature-based \\
    
    \addlinespace
    
    \citet{sharchilev2018startupsuccess} & Series A+ within 1 yr & -- & -- & 62.6\% & Feature-based \\
    
    \addlinespace
    
    \citet{gavrilenko2023improvingstartupsuccess} & Raise Series A+ within 1 yr & -- & 82.7\% & 74.4\% & Feature-based \\
    
    \addlinespace
    
    \citet{maarouf2025fused} & IPO, AC, or funding & 74.3\% & 78.3\% & 59.8\% & Feature-based \\
    
    \addlinespace
    
    \citet{ozince2024vcfounder} & IPO/AC/funding $>$\$500M & 66.7\% & 64.7\% & 68.8\% & Persona-based \\
    
    \addlinespace
    
    \citet{xiong2023founderideafit} & N/A & \multicolumn{3}{c}{No backtesting} & Pro/contra arguments \\
    
    \addlinespace
    
    \citet{xiong2024gptree} & IPO/AC/funding $>$\$500M & 87.6\% & 27.1\% & 37.3\% & No mention \\
    
    \addlinespace
    
    \citet{corea2021hacking} & IPO, AC, or funding & -- & -- & -- & Feature-based \\
    \bottomrule
    \end{tabular}
    \caption{Comparison of startup success prediction studies: success criteria, predictive performance of the best model, and interpretability.}
    \label{tab:related_work_overview}
\end{table*}

\end{document}